\documentclass[submission, PhysCore]{SciPost}

\hypersetup{
    colorlinks,
    linkcolor={red!50!black},
    citecolor={blue!50!black},
    urlcolor={blue!80!black}
}

\usepackage[bitstream-charter]{mathdesign}
\usepackage{enumerate}
\usepackage{cite}
\usepackage{xspace}
\usepackage{ifthen}
\usepackage{url}
\usepackage{booktabs}
\usepackage{multirow}
\usepackage{subcaption}
\usepackage{marginnote}
\usepackage{lineno}
\usepackage{amsmath,bm,graphicx}
\usepackage[utf8]{inputenc}

\newcommand{\pythia}{P\protect\scalebox{0.8}{YTHIA}\xspace}

\newcommand{\herwig}{\protect\scalebox{0.8}{HERWIG}\xspace}
\newcommand{\pytppp}{P\protect\scalebox{0.8}{YTHIA}8\xspace}

\newcommand{\atlas}{ATLAS\xspace}

\newcommand{\angantyr}{Angantyr\xspace}

\newcommand{\rivet}{Rivet\xspace}

\renewcommand{\eqref}[1]{eq.~(\ref{#1})\xspace}

\newcommand{\fig}[1]{\ref{#1}}
\newcommand{\figref}[1]{figure~\fig{#1}}

\newcommand{\citeref}[1]{ref.~\cite{#1}}
\newcommand{\citerefs}[1]{refs.~\cite{#1}}

\newcommand{\sect}[1]{\ref{#1}}
\newcommand{\sectref}[1]{section~\sect{#1}}

\newcommand{\appref}[1]{appendix~\sect{#1}}

\def\text{\mathrm}
\def\eg{\emph{e.g.}\xspace}
\def\ie{\emph{i.e.}\xspace}

\def\mrm#1{\mathrm{#1}}

\def\tauH{\ensuremath{\tau_{H}}}

\def\pp{\ensuremath{\mrm{pp}}\xspace}
\def\ppbold{\ensuremath{\mrm{\bf pp}}\xspace}

\def\pA{\ensuremath{\mrm{p}A}\xspace}

\def\ee{\ensuremath{\mrm{e}^+\mrm{e}^-}}

\def\AA{\ensuremath{AA}\xspace}

\def\keff{\ensuremath{\kappa_{\mrm{eff}}}}

\def\ColText(#1,#2)[#3]#4{\Text(#1,#2)[#3]{#4}}

\def\showcommentsflag{0}
\newcommand{\showcomments}{\def\showcommentsflag{1}}

\newcounter{commentcounter}%
\definecolor{armygreen}{rgb}{0.29, 0.33, 0.13}
\newcommand{\Red}[1]{\textcolor{red}{#1}}
\newcommand{\Black}[1]{\textcolor{black}{#1}}

\newcommand{\comment}[1]{\ifnum\showcommentsflag > 0%
	\addtocounter{commentcounter}{1}%
	{{\Red{\ensuremath{\ddagger^{\arabic{commentcounter}}}}\Black{}}}%
	\marginpar{\raggedright\tiny\it{{\Red{\ensuremath{\ddagger^{\arabic{commentcounter}}}}} {#1}\Black{}}}
	\fi%
}
\newcommand{\commentdel}[2]{\ifnum\showcommentsflag > 0%
	\Red{\sout{#1}}\comment{#2}%
	\fi
}
\newcommand{\commentadd}[2]{\ifnum\showcommentsflag > 0%
	\comment{#2}\Red{#1}%
	\else
	#1
	\fi
}
\newcommand{\commentchange}[3]{\ifnum\showcommentsflag > 0%
	\Red{\sout{#2}}\comment{#3}\Red{#1}%
	\else
	#1
	\fi
}
\newcommand{\nocomment}[1]{\ifnum\showcommentsflag > 0%
	{\tiny\it\Red{\{#1}\}}
	\fi%
}
\newcommand{\nocommentdel}[1]{\ifnum\showcommentsflag > 0%
	\Red{\sout{#1}}%
	\fi
}
\newcommand{\nocommentadd}[1]{\ifnum\showcommentsflag > 0%
	\Red{#1}%
	\else
	#1
	\fi
}
\newcommand{\nocommentchange}[2]{\ifnum\showcommentsflag > 0%
	\Red{\sout{#2}}\Red{#1}%
	\else
	#1
	\fi
}

\showcomments

\makeatletter
\renewcommand\paragraph{\@startsection{paragraph}{4}{\z@}%
  {-3.25ex\@plus -1ex \@minus -.2ex}%
  {1.5ex \@plus .2ex}%
  {\normalfont\normalsize\bfseries}}
\makeatother

\DeclareSymbolFont{usualmathcal}{OMS}{cmsy}{m}{n}
\DeclareSymbolFontAlphabet{\mathcal}{usualmathcal}

\graphicspath{{../old-paper/}}

\begin{document}

\begin{center}{\Large \textbf{
Jet modifications from colour rope formation in dense systems of non-parallel strings\\
}}\end{center}

\begin{center}
Christian Bierlich, Smita Chakraborty\textsuperscript{$\star$}, G\"{o}sta Gustafson, and Leif L\"{o}nnblad
\end{center}

\begin{center}
Department of Astronomy and Theoretical Physics, S\"{o}lvegatan 14A, S-223 62 Lund, Sweden
\\

${}^\star$ {\small \sf smita.chakraborty@thep.lu.se},\\
$\dagger$ MCNET-22-02, LU-TP 22-09
\end{center}

\begin{center}
\today
\end{center}


\section*{Abstract}

\textbf{We revisit our rope model for string fragmentation that has
  been shown to give a reasonable description of strangeness and
  baryon enhancement in high-multiplicity \ppbold events at the LHC. A
  key feature of the model is that the enhancement is driven by the
  increased string tension due to strings overlapping in dense
  systems. By introducing an improved space--time picture for the
  overlap between fragmenting strings, where also non-parallel strings
  are properly taken into account, we are now able to investigate the
  enhancement both in jets and in the underlying event in a consistent
  way.  }

\vspace{10pt}
\noindent\rule{\textwidth}{1pt}
\tableofcontents\thispagestyle{fancy}
\noindent\rule{\textwidth}{1pt}
\vspace{10pt}

\section{Introduction}
\label{sec:intro}

One of the most characteristic features of Quark--Gluon Plasma (QGP) formation in heavy ion (AA) collisions, is that of so--called ``jet quenching'' \cite{PHENIX:2004vcz}. In heavy ion collisions, jet quenching is mainly seen in energy loss or dispersion effects, manifest as, for example, suppression of high $p_\perp$ particle yields, with respect to scaled proton--proton (\pp) case \cite{STAR:2003pjh} or the suppression of away-side jets in central collisions \cite{STAR:2002svs}. With the higher energies available at LHC, the phenomenon has also been explored using 
$\mathrm{Z}$ bosons plus jets, where the $\mathrm{Z}$ decaying to leptons is used as an unaffected probe, to gauge the effect on the jet traversing the QGP \cite{CMS:2017eqd}.

Several experimental signatures for QGP production have, however, also
been observed in high multiplicity pp collisions, including
strangeness enhancement \cite{ALICE:2016fzo} and long-range
multi-particle correlations \cite{CMS:2016fnw}, more commonly known as
``collective flow''. Jet quenching effects have so far not been
observed in small systems (pp or proton--ion, \pA), which begs the
question if jet modification phenomena are completely absent in small
systems, or if the correct way to look for it has just not been
established \cite{Adolfsson:2020dhm}. One obvious reason for the
difficulty, is that it is not possible, like in \AA\
collisions, to look at differences when comparing to
similar measurements for pp collisions. Comparisons
with theoretical expectations are also difficult, as the expected
effects from quenching in small systems are very small, and the signal
is strongly affected by (uncertain) effects from initial state
radiation. It should also be mentioned that most theoretical
descriptions of jet quenching assumes that the jet is formed in a
deconfined ``bath'' of free partons (\ie the QGP), which may not be
appropriate in small systems, where at most a few droplets would
form. This includes approaches such as QGP-modified splitting kernels
\cite{Majumder:2009zu} at high virtualities, coupled with shower
modifications by transport theory \cite{Schenke:2009gb,He:2015pra} at
lower ones, but also approaches like the one offered by JEWEL
\cite{Zapp:2013vla}, where partonic rescattering off medium partons
are combined with the Landau--Pomeranchuk--Migdal effect
\cite{Zapp:2012ak}.

It should be noted that there are other mechanisms that may influence
jet production. In particular, some generators include ``colour
  reconnections'', especially in combination with multi-parton
interactions (see, \eg, \cite{Sjostrand:1987su,Gieseke:2012ft}), but
it has been shown (\eg, in \cite{Gras:2017jty} and
\cite{Chahal:2022rid}), that colour reconnections may influence jet
shapes even in $e^+e^-$ annihilations. In addition it has been shown
that the baryonic colour reconnection model in \herwig
\cite{Gieseke:2017clv} can give rise to strangeness enhancement in
dense environments in \pp\ collisions.


In a series of papers we have demonstrated, that collective flow and enhancement of strangeness and baryons, can be reproduced in high multiplicity pp events as a result of string-string interaction, when the infinitely thin string is generalized to a confining colour fluxtube, similar to a vortex line in a superconductor \cite{Baker:1991bc}. As discussed in \citeref{Bierlich:2019ixq}, models of string interactions offers a novel and convenient framework for studying jet modifications in small systems, as they are implemented in the general purpose Monte Carlo event generator \pythia, which allows the user to generate realistic collision events, with the effects switched ``on'' or ``off''. The study of jet modification effects does therefore not need to rely on a (non-existing) reference system.

The aim of this paper is therefore to look at possible effects of jet modification via increased strangeness and baryon numbers in jets. A very important tool is here the method developed in \citeref{Bierlich:2020naj}, to account for the interaction between strings which are not parallel to each other. This  was not possible in earlier versions of string-string interaction, but is naturally very important for handling the interaction between string pieces connected to a jet and strings in the underlying event.

The remainder of this paper is organised as follows. In
\sectref{sec:strings}, we recap the Lund string hadronization
framework, taking into account the transverse extension of strings, and
discuss how the string tension increase when such strings overlap,
leading to strangeness and baryon enhancement. Then we present the
parallel frame and our updated rope model in
\sectref{sec:parallel-frame-and-ropes}. In \sectref{sec:results} we
investigate how the average string tension varies as a function of
multiplicity and transverse momentum, and then investigate the
observable modifications the updated rope model predicts for jets and
the underlying event in \pp\ collisions at the LHC, before we present
our conclusions in \sectref{sec:conclusion}.

\section{String hadronization and colour fluxtubes}
\label{sec:strings}

In this section we will briefly introduce relevant parts of the Lund string hadronization
model, building up to the rope hadronization model used for the model results. For more detailed
reviews on Lund strings, we refer the reader to the large body of existing literature.
The original papers deal mainly with hadronization of a single
straight string \cite{Andersson:1978vj,Andersson:1983jt}. Gluons were introduced as `kinks` on a string in refs. \cite{Andersson:1979ij,Andersson:1980vj}. Somewhat dated reviews are presented in refs. \cite{Andersson:1983ia,Andersson:1997xwk}, and a number of recent papers on Lund strings
present the model in a more modern context \cite{Ferreres-Sole:2018vgo,Duncan:2019poz,Bierlich:2020naj,Bierlich:2022vdf}, including our original paper on rope hadronization \cite{Bierlich:2014xba}.

The Lund string is a "massless relativistic string" (or a "Nambu-Goto string"). Such a string has no transverse extension, and it also has no longitudinal momentum, which implies that it is boost invariant. \footnote{For the kinematics of such a string see \citeref{Artru:1979ye}.} 
This may be a good approximation for a linear colour fluxtube, where the width is not important.
In \sectref{sec:fluxtubes} we will
discuss going beyond this approximation.

\subsection{Lund string hadronization}
\label{sec:hadronization}

\paragraph{Hadronization of a straight string}

We first look at a single, straight string stretched between a quark and an
anti-quark. 
The string can break via $q\bar{q}$ pair creation, 
in a process which can be regarded as a tunneling process as discussed in \citeref{Brezin:1970xf}. For a single quark species the production probability is given by
\begin{equation}
	\label{eq:tunnel}
	\frac{\mrm{d}\mathcal{P}}{\mrm{d}^2p_\perp} \propto \kappa \exp\left(-\frac{\pi \mu^2_\perp}{\kappa}\right).
\end{equation}
Here $\mu^2_\perp = \mu^2 + p^2_\perp$ is the $\textit{quark}$ squared transverse mass. The exponential conveniently 
factorizes, leaving separate expressions for selection of mass and $p_\perp$ to be used in the Monte Carlo event generator. 
 With $\kappa \approx$ 1 GeV/fm, this result implies that strange quarks are suppressed by 
roughly a factor 0.3 relative to a u- or a d-quark (and that the probability to produce a c-quark with 
this mechanism is $\sim 10^{-11}$).
It also means that the quarks are produced with an average $p_\perp \sim 250$ MeV, 
independent of its flavour.

When the quarks and antiquarks from neighbouring breakups combine to mesons, their momenta can be calculated as an iterative process. The hadrons are here ``peeled off''
one at a time, each taking a fraction ($z$) of the \textit{remaining} 
light-cone momentum ($p^\pm = E \pm p_z$) along the positive or negative
light-cone respectively. The probability for a given $z$-value is here given by 
\begin{equation}
\label{eq:lu-frag}
f(z) \propto \frac{(1-z)^a}{z}\exp(-bm_\perp^2/z).
\end{equation}
Here $m_\perp$ is the transverse mass of the \textit{meson}, and the two parameters $a$ and $b$ are to be determined by tuning to data from
$\mrm{e}^+\mrm{e}^-$ collisions.
In principle the $a$--parameter could depend on the quark species, but in default \pythia (the Monash tune) it is the same for strange and non-strange quarks. Baryon--antibaryon pairs can be produced via production of a diquark--antidiquark pair, and in this case the $a$-parameter has to be modified. The parameter $b$ must, however, be universal.

An important consequence of \eqref{eq:lu-frag} is the probability
distribution in proper time ($\tau$) for string breakup vertices. 
Expressed in terms of the quantity $\Gamma = (\kappa \tau)^2$,
the distribution is given by:

\begin{equation}
\label{eq:frag-time-dist}
	\mathcal{P}(\Gamma)\mrm{d}\Gamma \propto \Gamma^a \exp(-b\Gamma) \mrm{d}\Gamma.
\end{equation}
The breakup-time is an important ingredient for string interactions, as the hadronization
time sets an upper limit on the available time for strings to push each other and form ropes.
As such, hadronization of a system of interacting strings will \textit{not} happen when the
system has reached equilibrium, but will be cut off when the string hadronizes. For strings
hadronizing early, one can then imagine a mixed phase of strings and hadrons, before the transition
to a pure hadron cascade. In this paper we consider only the effect of string interactions, and
leave the interplay with the hadronic cascade for a future paper. We note, however, that a full
hadronic cascade has recently been implemented in \pythia \cite{Sjostrand:2020gyg,Bierlich:2021poz}, revealing only minor 
effects in proton collisions.
Typical values for $a$, $b$ and $\kappa$ give an average breakup time of around 1.5 fm. 
This can not be identified as the hadronization time (or freeze-out time). This could equally  
well be interpreted as the time when the quark and the antiquark meet for the first time. 
In addition the breakup times fluctuate, and each string will hadronize at different times.

\paragraph{Gluons and non-straight strings}
\label{sec:kinkygluon}

An essential component in the Lund hadronization model is that a gluon 
is treated as a point-like ``kink'' on the string, carrying energy and 
momentum. 
A gluon carries both 
colour and anti-colour, and the string can be stretched from a quark, via a set 
of colour-ordered gluons, to an anti-quark (or alternatively in a closed 
loop of colour-ordered gluons).

When a gluon has lost its energy, 
the momentum-carrying kink is split in two corners, moving with the 
speed of light but carrying no momentum, stretching a new straight string 
piece between them. When two such corners meet, they can “bounce off”; 
the string connecting them then disappears, but a new one is “born”.
In a pp collision a typical string will contain several gluons, connected by 
string pieces which are stretched out, may disappear and then be replaced by 
new string pieces.
All these string pieces move transversely in different directions, but at any time the string consists of a set of straight pieces. For a 
description of how such a string hadronizes, we refer to 
\citerefs{Sjostrand:1984ic, Bierlich:2020naj}. The interaction between strings with several non-parallel pieces is discussed in \sectref{sec:parallel-frame-and-ropes}.

\subsection{Strings as colour fluxtubes}
\label{sec:fluxtubes}

The description of a confining colour field by an infinitely thin string is necessarily an approximation, 
relevant only when the result is insensitive to the width.
In high multiplicity events this is no longer the case, and the strings have to be 
treated as colour fluxtubes, with a non-zero width.
We here first discuss the properties of a single fluxtube, and then the interaction between two or more parallel fluxtubes. The generalization to non-parallel fluxtubes is presented in 
\sectref{sec:parallel-frame-and-ropes}.

\subsubsection{A single fluxtube}

The simplest model for a QCD fluxtube is the MIT 
bag model \cite{Johnson:1975zp}. 
Here a homogenous longitudinal colour-electric field is kept inside a tube by the pressure from the vacuum condensate.
An improved description is obtained in lattice calculations.
A common method is here to use the method of Abelian projections, proposed by 't\,Hooft \cite{tHooft:1981bkw},
 which is based on partial gauge fixing. The result of these calculations
show that the field is dominated by a longitudinal colour-electric
field, surrounded by a transverse colour-magnetic current in the
confining vacuum condensate \cite{Nishino:2019bzb, Shibata:2019ghh}.
This picture is very similar to the confinement of the magnetic field
in a vortex line in a superconductor (with electric and magnetic
fields interchanged, see \textit{e.g.} \citeref{Kondo:2014sta}).

As observed in \citeref{Bierlich:2020naj} the
measured shape of the colour electric field obtained in \citeref{Cea:2015wjd} is well approximated by a
Gaussian distribution:
\begin{equation}
  \label{eq:Gaussian}
  E(\rho) = E_0  \exp \left( -\rho^2 /2R^2 \right),
\end{equation}
where $\rho$ is the transverse distance in cylinder coordinates.
The width of a fluxtube is difficult to estimate in lattice
calculations, as it is naturally given in lattice units, see
\textit{e.g.} \citeref{Sommer:2014mea}. In the bag model, the width is roughly given by
$\sqrt{\langle \rho^2 \rangle} \approx 1$~fm (where $\rho$ is the radial
coordinate), while lattice calculations give radii of $ \sim 0.5$~fm
or even less \cite{Cea:2014uja,Baker:2018mhw, Nishino:2019bzb}.  

The field density in \eqref{eq:Gaussian}  is related to the string tension through
\begin{equation}
  \label{eq:gausstension}
  \int d^2\rho E^2(\rho)/2=\pi E_0^2 R^2 = g\kappa,
\end{equation}
where $g$ is the fraction of the total energy of the string associated
with the colour electric field. We expect $g$ to be of the order 1/2, which is 
the value obtained in the bag model, where the energy in the field and the expelled condensate are of equal size.
For a further discussion of the vacuum condensate and colour fluxtubes we refer 
to \citeref{Bierlich:2017vhg} and references therein.

\subsubsection{Interacting parallel fluxtubes}

High multiplicity collisions will give a high density of fluxtubes,
with a corresponding high energy density. 
In \citeref{Bierlich:2020naj} we discussed the collective
effects expected from the initial expansion, and in this paper we will
concentrate on the effects of rope hadronization, and in particular
study the production of strange hadrons.
Here we first restate our  treatment of interaction between parallel fluxtubes, presented in 
\citeref{Bierlich:2014xba}. How this can be generalized to a general situation with non-parallel fluxtubes will be discussed in \sectref{sec:parallel-frame-and-ropes} below.

\paragraph{Rope formation}

For two overlapping parallel fluxtubes, separated by a transverse distance $\delta$,
we get from \eqref{eq:Gaussian} the interaction energy of the field
\begin{equation}
  \label{eq:int-energy}
  \int d^2\rho (\mathbf{E}_1(\rho)+\mathbf{E}_2(\rho))^2/2 - 2\int d^2\rho E^2(\rho)/2=
  \int d^2\rho \mathbf{E}_1(\rho)\cdot\mathbf{E}_2(\rho)=
  2 \pi E_0^2 R^2 e^{-\delta^2/4R^2}.
\end{equation}
Such a system will expand transversely, and if it does not hadronize before, it will reach
equilibrium, where the energy density corresponds to the free energy
density in the vacuum condensate. 

The expression in \eqref{eq:int-energy} does not include
the surface energy for the combined flux tube. In the bag model this is zero, and in equilibrium the transverse area will be doubled, and the interaction energy will be zero. 
For a vortex line in a dual QCD superconductor, it depends on the properties of the superconductor, but also here the interaction energy
will be much reduced at the time of hadronization.
It will then be necessary to go beyond the Abelian approximation. 
For two fluxtubes stretched by quarks, the two quarks can either form a colour sextet or an anti-triplet, and
with more fluxtubes also higher multiplets are possible.
Here lattice calculations show that
a set of overlapping strings form a "rope", with a tension proportional to the second Casimir operator for the colour multiplet at the end of the rope \cite{Bali:2000un}.

Biro, Nielsen, and Knoll pointed out \cite{Biro:1984cf} that if a rope
is formed by a number of strings with random charges, they add up as a random 
walk in colour space. This implies that the net colour grows as the square root of the 
number of strings. A rope stretched by $m$ colour charges and $n$ anti-charges can
then form a colour multiplet characterised by two
numbers $p$ and $q$, such that an arbitrary state, by a rotation in colour space,
can be transformed into a state with $p$ coherent colours (\emph{e.g.} red) and $q$
coherent anti-colours (\emph{e.g.} anti-blue), such that the colour and the anti-colour 
do not form a colour singlet. Such a multiplet is denoted $\{p,q\}$, and we always have $p\leq m$ and $q \leq n$.

For any such multiplet we can write down the number of states, \ie the 
multiplicity\footnote{The multiplicity provides the standard nomenclature for multiplets, where $N = 1$ is called ``singlet'', $N=3$ is called ``triplet'', $N=6$ is called ``sextet'' etc. We will here, when necessary, use the slightly more verbose notation $\{p,q\}$,
which allows one to distinguish between \eg a triplet and an anti-triplet.} of the multiplet:
\begin{equation}
\label{eq:SU3-colour-conf}
N = \frac{1}{2} (p+1) (q+1) (p+q+2).
\end{equation}
As mentioned above, the total
tension of such a rope is proportional to the second Casimir operator for the
multiplet, which gives
\begin{equation}
\label{eq:casimir-operator}
\kappa^{\{p,q\}}=\frac{C_2({p,q})}{C_2({1,0})}\kappa^{\{1,0\}} =\frac{1}{4} \left(p^2 + pq + q^2 + 3p + 3q\right) \kappa^{\{1,0\}},
\end{equation}
where $\kappa^{\{1,0\}}\equiv\kappa$ is the tension in
a single string.

In the \pythia treatment used here, there are, however, other effects also addressing 
string coherence effects. Importantly, parts of this colour summation is in an approximate
way treated by “colour reconnection”.  As a simple example we can look
at two anti-parallel strings with triplet--anti-triplet pairs in each
end. These can either form an octet or a singlet, with probabilities
8/9 and 1/9 respectively. Here the octet (denoted $\{1,1\}$) gives
\begin{equation}
  \kappa^{\{1,1\}}=\kappa\cdot C_2^{\{1,1\}}/C_2^{\{1,0\}}=9\kappa/4.
  \label{eq:octetkappa}
\end{equation}
 The singlet ($\{0,0\}$), with no string at all, gives $\kappa^{\{0,0\}}=0$.

The colour reconnection process in a situation with several strings
can be related to an expansion in powers of $1/N_c$, as discussed in
\citerefs{Lonnblad:1995yk, Christiansen:2015yqa}.

For the special case of $N_c=3$ there is also a different kind of
reconnection. For a rope formed by two parallel strings, the two
triplets in one end can give either a sextet or an anti-triplet (and a
corresponding anti-sextet or triplet in the other end) with
probabilities 2/3 and 1/3 respectively. For the latter we simply have
just a single string.

The two original colour triplets are connected in a “junction”, and
such a reconnection can be particularly important for baryon
production.  This possibility is not implemented in the present
version of our Monte Carlo, but will be included in future work.  We
note that for an arbitrary number of colours, the corresponding
situation is only obtained when $N_c - 1$ colour charges combine to
one anti-colour charge. The junction formation with three strings does
therefore, for $N_c \neq 3$, correspond to a configuration where $N_c$
strings are connected, which cannot be directly interpreted as a
$1/N_c$ correction.

We will in the following adopt a picture where the process of string (rope)
fragmentation follows after a process of colour reconnections, and that
this will leave the system in a state with $p$ parallel and $q$ anti-parallel
strings forming a coherent multiplet $\{p,q\}$. 

\paragraph{Rope hadronization}

A rope specified by the multiplet $\{p,q\}$, can 
break via a succession of single  $q\bar{q}$ productions,
through the tunnelling mechanism in \eqref{eq:tunnel}. In each step a multiplet
$\{p,q\}$ is changed to either $\{p-1,q\}$ or $\{p,q-1\}$.
It is here important to note that
\emph{the tunneling is not determined by the total tension in the rope, but by the 
energy released, determined by the
reduction in the tension}
 caused by the production of the new $q\bar{q}$ pair. 
Hence, we get from \eqref{eq:casimir-operator} an effective string tension, when 
the field goes from $\{p+1,q\}$ to $\{p,q\}$, given by
\begin{equation}
\label{eq:eff-kappa}
\keff=\kappa^{\{p+1,q\}}-\kappa^{\{p,q\}}=\frac{2p+q+4}{4} \kappa.
\end{equation}

The consequence of this picture is that we can treat the rope
fragmentation as the sequential decay of the individual strings
forming the rope, much in the same way as an everyday rope would break
thread by thread. Technically it means that we can use the normal
string fragmentation procedure in \pytppp, with the modification that
we in each break-up change the fragmentation parameters according to
the effective string tension calculated from the overlap of
neighbouring strings.  The changes to these parameter explained in
detail in \citeref{Bierlich:2014xba}, and are for reference also
listed in \appref{sec:kappaeffects}. The changes are somewhat
convoluted, since most of the parameters only indirectly depend on
the string tension, but the main effect easily seen in
\eqref{eq:tunnel}, namely that an increased string tension will
increase the probability of strange quarks and diquarks relative to
light quarks in the string breakup.

\section{Rope hadronization with non-parallel strings}
\label{sec:parallel-frame-and-ropes}

Our previous work on rope formation \cite{Bierlich:2014xba} relied on
the assumption that strings in high energy hadron collisions can be
assumed to be approximately parallel to each other and to the beam
axes. This prevented a detailed investigation of possible effects in
hard jets, especially those traversing the dense environment of an
\AA\ collision. In our recent work on the shoving model
\cite{Bierlich:2020naj} we found a remedy where the interaction
between any pair of strings can be studied in a special Lorentz frame,
even if they are not parallel to each other or to the beam.  We call
it ``the parallel frame'', and it can be shown that any pair of straight
string pieces can be transformed into such a frame, where they will
always lie in parallel planes.

Below we will use this parallel frame to calculate the increased
string tension in the rope formation of arbitrarily complex string
configurations.

\subsection{The \textit{parallel frame} formalism}
\label{sec:parallel-frame}

In the previous rope implementation \cite{Bierlich:2014xba}, the way
to determine if any two string pieces are overlapping was to boost
them to their common centre-of-mass frame and here measure the distance
between them at a given space-time point of break-up. This was
done in a fairly crude way, not really taking into account that the
two string pieces typically cannot be considered to be parallel in
this frame. In general there is no frame where two arbitrary
string pieces can be considered to be exactly parallel, but in the
\textit{parallel frame} introduced in \citeref{Bierlich:2020naj} it can
be shown that any two string pieces will always be stretched out in
parallel planes in a symmetric way. This works for all pairs of string pieces, even if
one piece is in a high transverse momentum jet and the other is in the
underlying event.

\begin{figure}
  \centering
  \includegraphics[width=0.45\textwidth,clip]{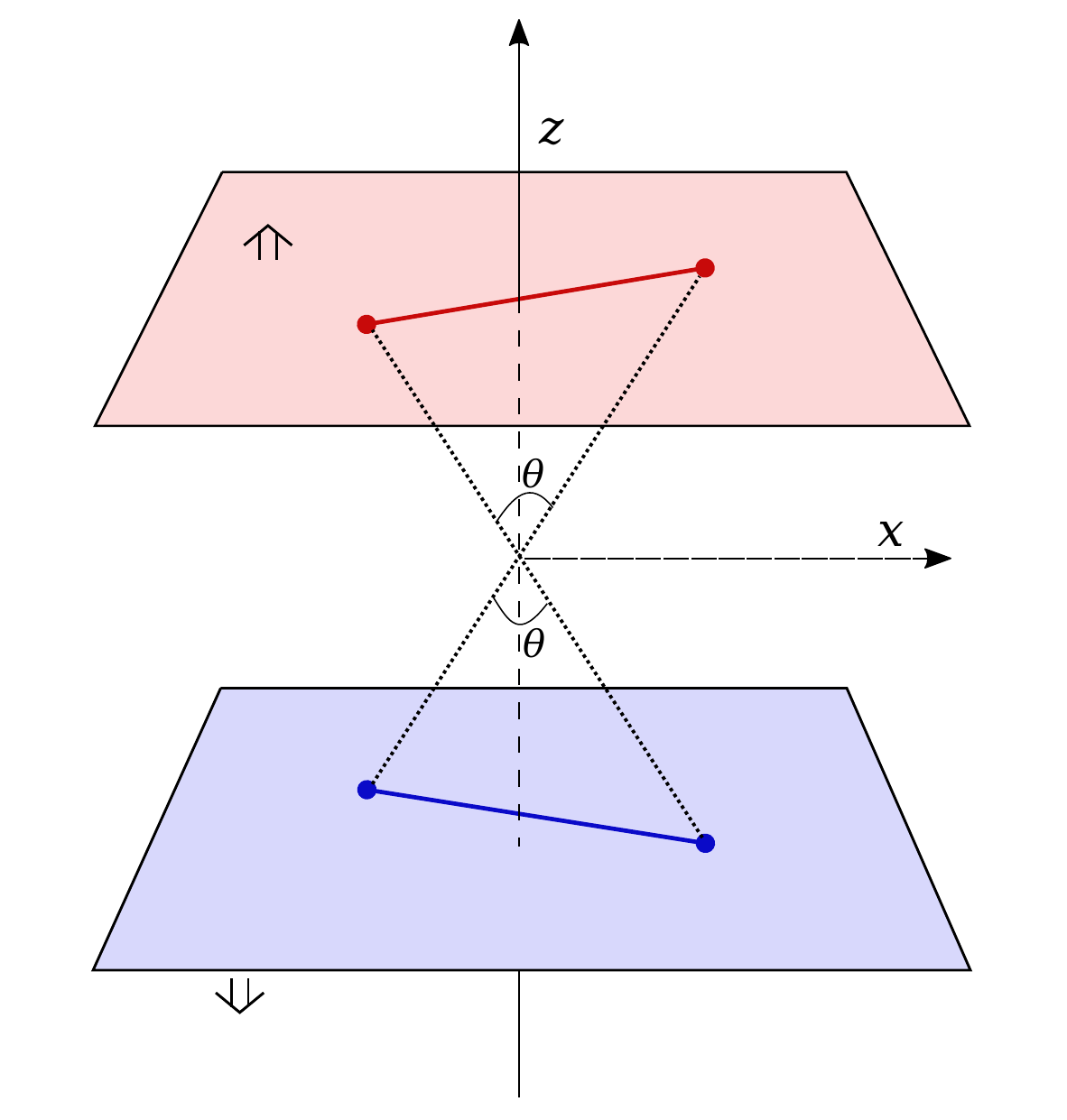}
  \includegraphics[width=0.45\textwidth,clip]{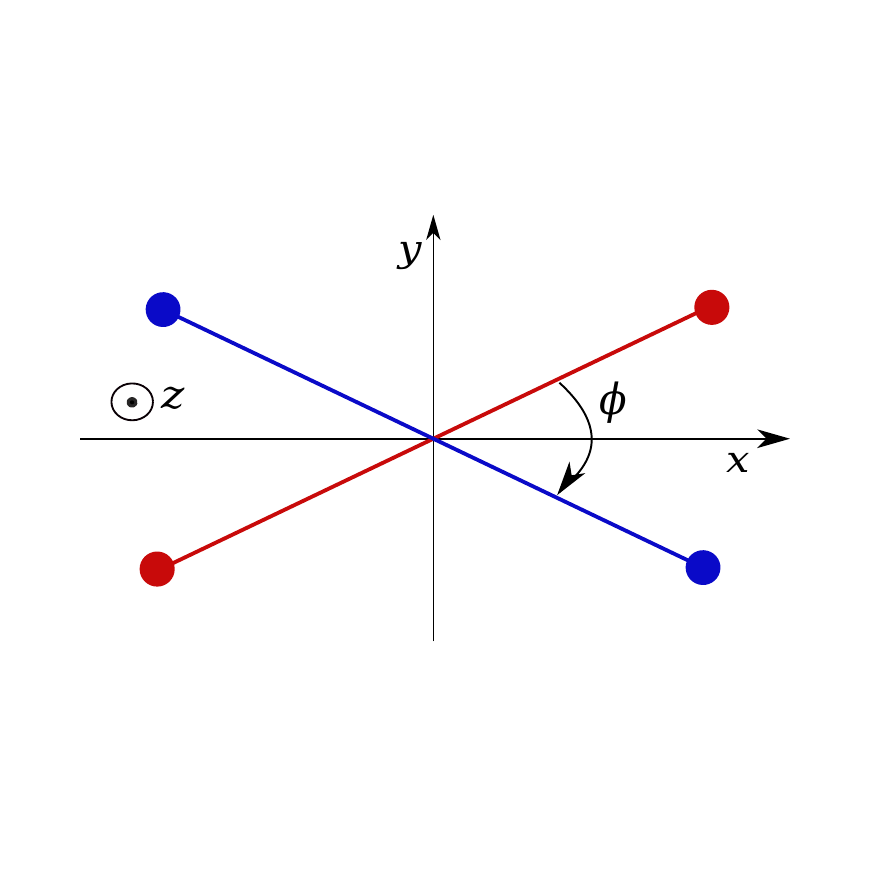}
  \caption{The parallel frame showing the parallel planes of two
    strings and the opening angle $\theta$ and skewness angle $\phi$.}
  \label{fig:parallelframe}
\end{figure}

In \figref{fig:parallelframe} we show a space--time picture of
two string pieces stretched between two pairs of partons in this
parallel frame. Since massless partons are propagating at the speed of
light irrespective of the magnitude of their momenta, only the
angles between them are important for the following.
In the parallel frame the two string pieces have the same
opening angle $\theta$, and the partons of one piece propagate with
an angle $\theta/2$ w.r.t.\ the $z$-axis. The partons of the
other propagate in the opposite direction, with an angle
$\pi-\theta/2$. At any given time, both string pieces will lie in
planes parallel to the $xy$-plane and to each other. Looking at the
projections of the string pieces on the $xy$-plane, we denote the
angle between them by $\phi$, and the frame is chosen such that all
partons form an angle $\phi/2$ with the $x$-axis.

\def\sinth{\ensuremath{\sin\frac{\theta}{2}}}
\def\costh{\ensuremath{\cos\frac{\theta}{2}}}
\def\sinph{\ensuremath{\sin\frac{\phi}{2}}}
\def\cosph{\ensuremath{\cos\frac{\phi}{2}}}
\def\sinhe{\ensuremath{\sinh\frac{\eta}{2}}}
\def\coshe{\ensuremath{\cosh\frac{\eta}{2}}}
To simplify the calculations we write the momenta of the partons using
their transverse momentum, $p_\perp$, and pseudo-rapidity difference, $\eta$, with respect to the $z$-axis, rather than the energy and opening polar angle
(where $p_z=e\costh=p_{\perp}\sinhe$), and get, using the notation
$p=(e;p_x,p_y,p_z)$,
\begin{eqnarray}
\label{eq:pT}
p_1 &=& p_{\perp1} \left( \coshe; \phantom{-}\cosph, \phantom{-}\sinph, \phantom{-}\sinhe \right), \nonumber \\
p_2 &=& p_{\perp2} \left( \coshe; -\cosph,    -\sinph, \phantom{-}\sinhe\right), \nonumber  \\
p_3 &=& p_{\perp3} \left( \coshe; \phantom{-}\cosph,  -\sinph, -\sinhe \right), \nonumber  \\
p_4 &=& p_{\perp4} \left( \coshe; -\cosph,  \phantom{-}\sinph, -\sinhe \right).
\end{eqnarray}
Clearly we have six degrees of freedom, and we can construct six
independent squared invariant masses, $s_{ij}=(p_i+p_j)^2$. This
means that for any set of four \emph{massless} partons we can (as
long as no two momenta are completely parallel) solve for
$p_{\perp i}$, which will give us:
\begin{equation}
\label{eq:transverse-momenta}
p^2_{\perp1} = \frac{s_{12}}{4} \sqrt{\frac{s_{13}s_{14}}{s_{23}s_{24}}},\,\,\,\,
p^2_{\perp2} = \frac{s_{12}}{4} \sqrt{\frac{s_{23}s_{24}}{s_{13}s_{14}}},\,\,\,\,
p^2_{\perp3} = \frac{s_{34}}{4} \sqrt{\frac{s_{13}s_{23}}{s_{14}s_{24}}},\,\,\,\,
p^2_{\perp4} = \frac{s_{34}}{4} \sqrt{\frac{s_{14}s_{24}}{s_{13}s_{23}}},
\end{equation}
and furthermore solve for the angles $\phi$ and $\eta$:
\begin{equation}
\label{eq:parallel-angles} 
\cosh\eta=\frac{s_{14}}{4p_{\perp 1}p_{\perp 4}}+\frac{s_{13}}{4p_{\perp 1}p_{\perp 3}} 
\quad\text{and}\quad
\cos\phi=\frac{s_{14}}{4p_{\perp 1}p_{\perp 4}}-\frac{s_{13}}{4p_{\perp 1}p_{\perp 3}}. 
\end{equation}
To further specify the frame we renumber the particles so that
$\phi<\pi/2$ to have the strings more parallel to the $x$-axis and not
to the $y$-axis, and we define the $x$-axis to be their combined
\emph{rope} axis.  The result is that for a breakup at a given
space--time point in one string piece, we can in the parallel frame
have a reasonable handle on the overlap with any other string piece.

\subsection{Overlap in the parallel frame}
\label{sec:overl-parall-frame}

In \eqref{eq:int-energy} we wrote down the interaction energy of two
completely parallel strings separated by a small distance. We now want
to use this to estimate the effective overlap of two strings that are
not completely parallel, but lie in parallel planes.

At a specific point along the $x$-axis in the parallel frame we denote
the separation between the strings in the $yz$-plane by
$(\delta_y,\delta_z)$ and integrate the interaction of the field given
the skewness angle $\phi$ to obtain
\def\csph{\cos\frac{\phi}{2}}
\def\cph{\cos\phi}
\def\ovl{\mathcal{I}(\delta_y,\delta_z,\phi)}
\begin{eqnarray}
  \ovl\!\! &=&  \!\!\int d^2\rho \mathbf{E}_1(\rho)\cdot\mathbf{E}_2(\rho)
               \nonumber\\
           &=& \!\!E_0^2\cph\!\!\int\!\!\! dy\,dz
           \exp\left(-\frac{y^2\csph+z^2}{2R^2}\right)
               \exp\left(-\frac{(y-\delta_y)^2\csph+(z-\delta_z)^2}{2R^2}\right)
               \nonumber\\
           &=& \!\!2\pi E_0^2R^2\frac{\cph}{\cosph}
               \exp\left(-\frac{\delta_y^2\csph+\delta_z^2}{4R^2}\right).
               \label{eq:overlap}
\end{eqnarray}
Here we note that the skewness angle enters both in the scalar product
and in the strength of the field along the $y$-axis, and that the
overlap vanishes for orthogonal strings.

We can now define the relative overlap as $\ovl/\mathcal{I}(0,0,0)$
and use it as a probability (assuming that
$\mathbf{E}_1\cdot\mathbf{E}_2>0$) that a breakup in one string is
affected by an increased string tension due to the overlap with the
other. This would then correspond to a $\{2,0\}\to\{1,0\}$ transition
giving an effective string tension $\keff=3\kappa/2$ in
\eqref{eq:eff-kappa}. If the strings instead points in the opposite
directions along the $x$-axis ($\mathbf{E}_1\cdot\mathbf{E}_2<0$) this
would correspond to a $\{1,1\}\to\{0,1\}$ breakup with
$\keff=5\kappa/4$.

In this way we can for each breakup in one string piece, take all
other string pieces in an event, and for each go to the parallel frame to
determine if it will contribute to $p$ or $q$. In our implementation described
below, we sum the relative overlaps in $p$ and $q$ respectively and
round them off to integers, rather than treating them as individual
probabilities for each pair of string pieces, which on average gives
the same result.

It should be pointed out that in the parallel frame we also have a
handle on which string breaks up first. If we assume that the string
breaks at a common average proper time along the string, \tauH, we can
in the parallel frame calculate the proper time of the other string in
space--time point where we calculate the overlap. If the latter is at
larger \tauH, we conclude that the other string has already broken up,
and can no longer contribute to an increased string tension in the
break-up being considered.

\subsection{Monte Carlo implementation}
\label{sec:monte-carlo-impl}

The main technical problem with implementing the rope model in
\pytppp, is the order in which the string fragmentation
proceeds. First, the flavour and transverse momentum of the break-up
is chosen (\eqref{eq:tunnel}) together with the type of the
chopped-off hadron. Only then is the momentum fraction, $z$, chosen
according to \eqref{eq:lu-frag}, and only then do we know exactly
where the string breaks and can calculate the \keff\ in that
point. But we need to know \keff\ to be able to calculate a break-up, so
we have a kind of \textit{Catch-22} situation.

The way we solve this is to perform a trial break-up to pre-sample the
overlap of a given string, and use the overlap there to get an approximate
\keff. Then we discard the sample break-up and produce a new one using
this \keff. On the average we will then get a reasonable estimate of
the overlap around a break-up. For a general break-up in the
underlying event this should be good enough, but if we are interested
in details of the hadron production in, \eg, the tip of a jet, this
procedure may be inappropriate (see further discussion below in
\sectref{sec:strangejet-results-pp}).

The procedure to calculate \keff\ looks as follows:\\[-8mm]
\begin{enumerate}\itemsep -1mm
\item Produce a trial break-up in the string being fragmented, and
  deduce from which string piece it comes.
\item Pair this piece with every other string piece in the event, make
  a Lorentz transformation to the parallel frame of each pair.
\item Using the pseudo-rapidity of the produced hadron in each such
  frame, and assuming the break-up occurred at the proper time, \tauH,
  find the space-time point of the break-up of the first string piece.
\item In the corresponding $yz$-plane determine the proper time of the
  other string piece and if that is less than \tauH, calculate the
  overlap according to \eqref{eq:overlap}, and determine if this
  overlap should contribute to $p$ or $q$ in the breakup.
\item With the summed $p$ and $q$ (rounded off to integer values), we
  now calculate \keff\ according to \eqref{eq:eff-kappa}.
\item Throw away the trial break-up with its produced hadron and
  change the \pytppp fragmentation parameters according to the
  obtained \keff\ and generate the final break-up.
\end{enumerate}

As mentioned in \sectref{sec:kinkygluon}, some care has to be taken
when it comes to soft gluons. Normally, all string pieces can be said
to be \textit{dipoles} between colour-connected partons, and in any
parallel frame this string piece is parallel to the $xy$-plane. But a
soft gluon may have lost all its momentum before the string breaks, and
the break-up can then occur in a piece of the string that is not
parallel to the string pieces of the connected dipoles. To include this
possibility we introduce \textit{secondary} dipoles, so that if we
have two dipoles connected to a soft gluon, \eg $q_i-g_j$ and
$g_j-\bar{q}_k$, then a secondary string will be included spanned
between the momenta of $q_i$ and $\bar{q}_k$, but using the space-time
point where the gluon has lost all its momentum to the connected
string pieces, as a point of origin.

The problem with soft gluons 
is present also for our shoving model
in \cite{Bierlich:2020naj}, and the solution with secondary dipoles is
now also used there. This will be described in more detail in a future
publication, where we also describe the procedure for including these
\textit{higher order} dipoles in cases where we have several
consecutive soft gluons along a string.

\subsection{Interplay with the Shoving model}
\label{sec:interpl-with-shov}

Clearly our rope model is very tightly connected with our shoving
model. They both rely on the parallel frame and technically they both
use the same infrastructure for looking at overlaps between string
pieces. However, here there is again a kind of
\textit{Catch-22}.

Physically the shoving precedes the hadronization, and pushes the
strings apart before they hadronize. As this affects the value of
\keff, the shoving should be executed first.  However, for technical
reasons the pushes are applied directly to the produced hadrons rather
than to the individual string pieces. Therefore we must calculate the
hadronization before we can execute the pushes.

We are currently working on a solution to this problem, and plan to
present it in a future publication. The main effect of the shoving is
expected to be a dilution of the strings resulting in a lowered
\keff. As discussed in \cite{Bierlich:2020naj} the precise value of
the string radius is not known, and in that paper we simply used a
canonical value of 1~fm. Also the string radius will affect the values
of \keff, and preliminary studies show that the effects of string
dilution from shoving are of the same order as moderate decrease of
the string radius of around 10\%.

\section{Results in \pp collisions}
\label{sec:results}

In this section, features of the rope hadronization model with the
parallel frame-formalism are investigated in \pp collisions. Since the
main feature of this new formalism is the much improved handling of
string pieces which are not parallel to the beam axis (\ie\ jets), we
will mostly concentrate on observables in events containing a process
with high momentum transfer, but in \sectref{sec:model-results} we
first show the behaviour in minimum bias collisions. Here the most
fundamental check of the dependence of \keff\ with final state
multiplicity, but more relevant for the parallel frame formalism, is
the dependence of \keff\ on particle $p_\perp$. In
\sectref{sec:jet-prop-results} we compare to existing experimental
results in the underlying event (UE) for $\mrm{Z}$-triggered
events. This is to ensure that the existing description of such
observables is not altered by our model. Finally in
\sectref{sec:strangejet-results-pp}, we present predictions for the
jet observables that are affected by rope formation in \pp.

\subsection{Model behaviour}
\label{sec:model-results}

\begin{figure}[t!]
  \centering
  \begin{minipage}{1.0\textwidth}
    \begin{center}
      \includegraphics[width=0.49\textwidth]{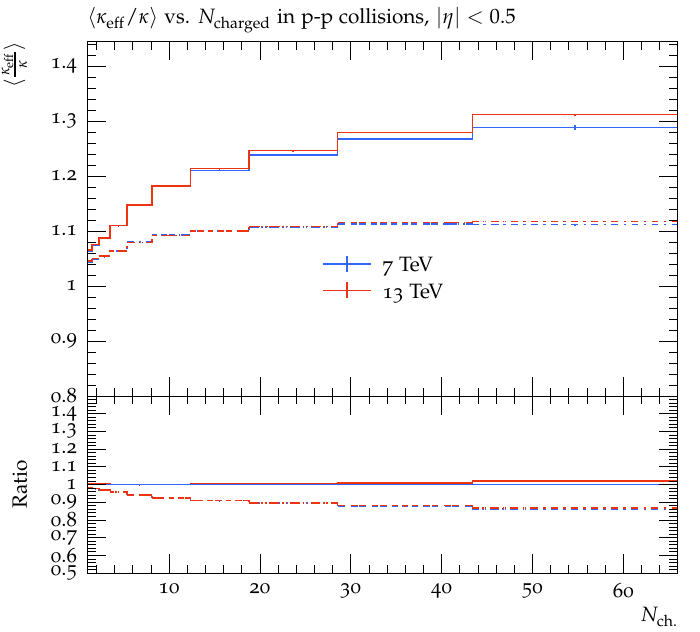}
      \includegraphics[width=0.49\textwidth]{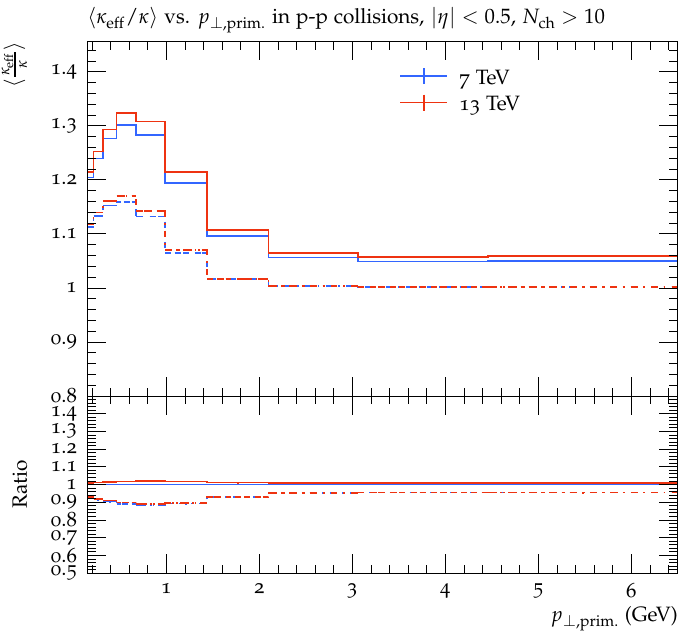}
    \end{center}
  \end{minipage}
  \caption{$\langle \keff/\kappa \rangle$ vs.\ $N_{\text{ch.}}$ (left)
    and vs.\ $ p_{\perp, \text{prim.}}$ for $N_{\text{ch.}} > 10$
    (right). Solid lines have string radius $R = 1$~fm and dot-dashed
    lines have $R = 0.5$~fm. Blue and red lines are for minimum-bias
    event at 7 and 13~TeV respectively.}
  \label{fig:kappa-mult-pT}
\end{figure}

In this section, we explore the variation of the effective string tension $\keff$ with rope hadronization, for minimum bias \pp events. The $\keff$ is shown for primary hadrons, \ie the effective string tension used to form a given hadron, produced directly in the hadronization process. Results are shown for two collision energies, $\sqrt{s} =$ 7 and 13~TeV, and two values of string radius, $R = 0.5$  and 1~fm.

In \figref{fig:kappa-mult-pT}, the dependence of $ \langle \keff/\kappa \rangle$ with respect to $ N_{\text{ch.}}$ in $|\eta| < 0.5$ is shown on the left, and $ p_{\perp, \text{prim.}}$ on the right. On the right, only events with $\mathrm{d}N_{\text{ch.}}/\mathrm{d}\eta>10$ are shown, to focus on events with several parton interactions. (At 13 TeV this corresponds to keeping roughly the 30\% of events with the highest multiplicity \cite{ALICE:2019dfi}.) 

On the left plot of \figref{fig:kappa-mult-pT}, it is seen that $\langle \keff/\kappa \rangle$ rises with around 30\% for $R = 1$~fm and 10\% for $R = 0.5$~fm, almost irrespective of $\sqrt{s}$, with the rise at 13 TeV being only slightly higher. The two main points to take away from this figure, is a) that $\mrm{d}N_\text{ch.}/\mrm{d}\eta$ is a good proxy for string density irrespective of collision energy, and thus works as a good scaling variable, and b) that any result will be very sensitive to the choice of $R$.
  
On the right plot of \figref{fig:kappa-mult-pT}, we observe that the increase in $\keff$ is larger for primary hadrons in the lower $p_\perp$ bins for both values of $R$. This means that the lower $p_\perp$ primary hadrons are formed from regions with high density of strings with more overlaps with adjacent strings. However, the higher $p_\perp$ partons correspond to ``mini-jet'' situations and are more separated in space-time from the bulk of strings. Such strings have less overlaps resulting in a lower $\keff$. Hence the high $p_\perp$ primary hadrons formed from such string break-ups show this effect.

In the lowest $p_\perp$ bins of $\langle \keff/\kappa \rangle$ vs $p_{\perp, \text{prim.}}$ plot, it is seen that $\keff$ drops to lower values. This behaviour arises from the fact that low $p_\perp$ particles are biased towards low $\keff$ values due to the $p_\perp$-dependence on $\kappa$ in the tunneling probability in \eqref{eq:tunnel}.

Overall we observe that rope hadronization significantly increases the
string tension at high-multiplicities and for low $p_\perp$
final-state particles. For higher $p_\perp$ the effect is smaller, but
does not disappear completely.

\subsection{Underlying event observables in reconstructed $\mrm{Z}$ events}
\label{sec:jet-prop-results}

\begin{figure}[t!]
	\centering
	\begin{minipage}{\textwidth}
		\includegraphics[width=0.498\textwidth]{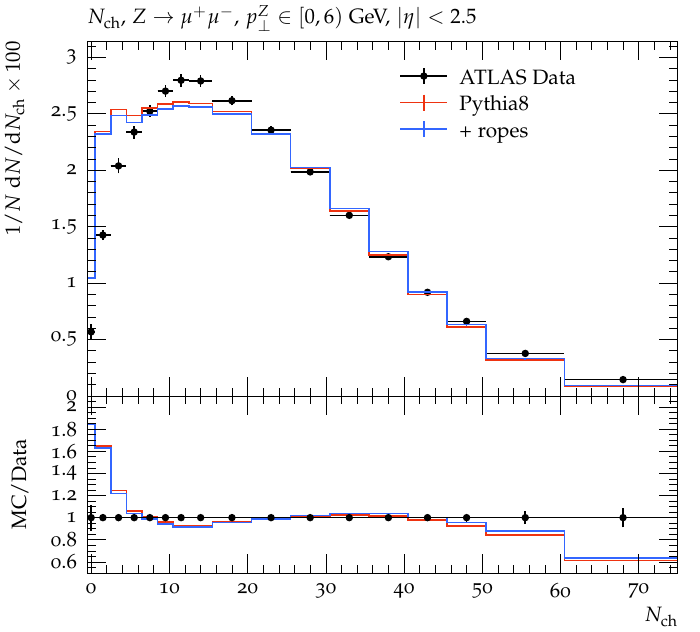}
		\includegraphics[width=0.498\textwidth]{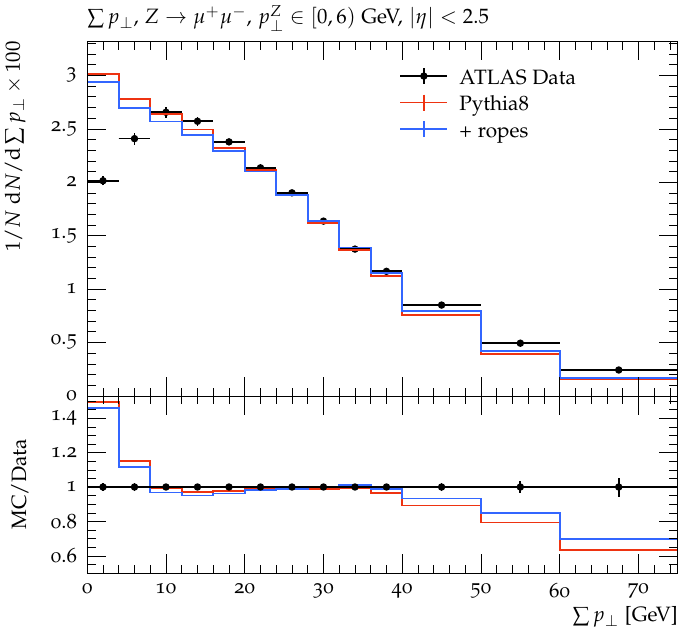}
		\includegraphics[width=0.498\textwidth]{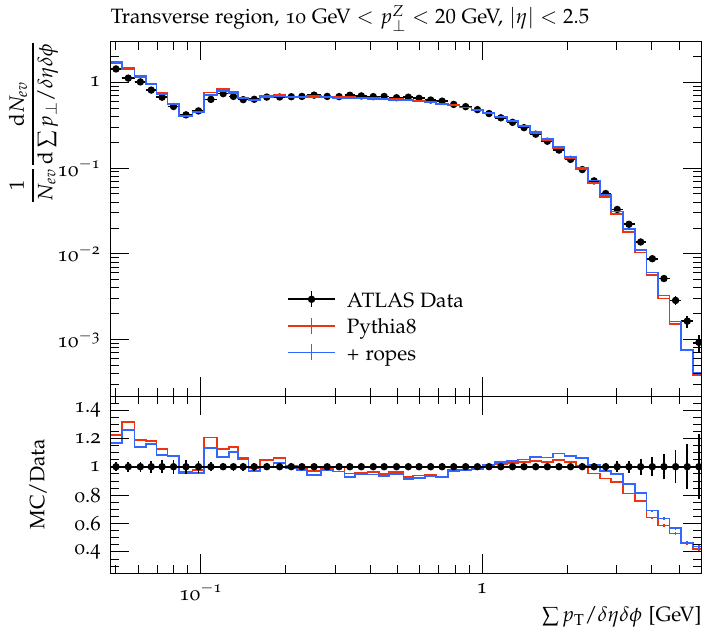}
		\includegraphics[width=0.498\textwidth]{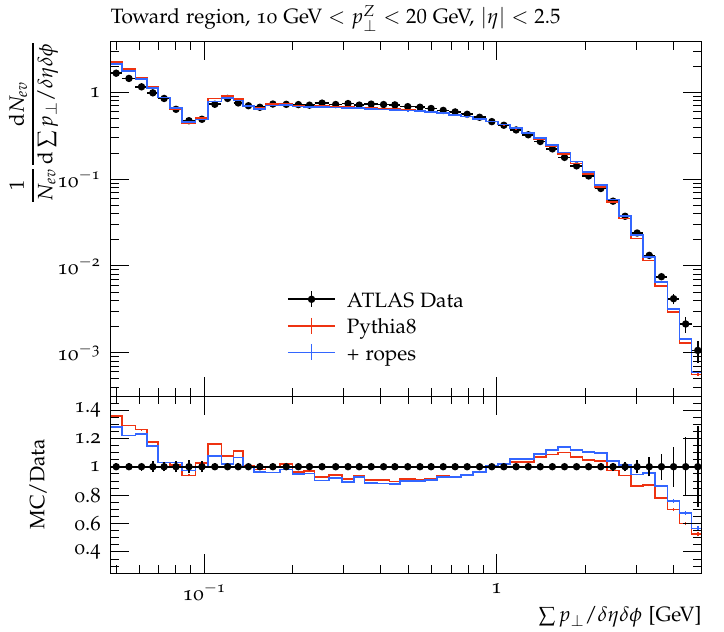}
	\end{minipage}
	\caption{Associated particle production in
		Z$\rightarrow \ell^-\ell^+$ events at $\sqrt{s}=7$~TeV compared to
		the default \pythia tune and with rope hadronization.  \textbf{Top
			row:} Distribution of charged particle multiplicity,
		$N_{\text{ch}}$, (top left) and summed scalar transverse momenta,
		$\Sigma p_\perp$ (top right) measured for events with $p_\perp^Z$
		range 0-6~GeV\cite{ATLAS:2014yqy}. \textbf{Bottom row:}
		$\Sigma p_\perp$ distributions in different azimuthal regions, in
		events with $p_\perp^Z$ range 10-20~GeV
		\cite{ATLAS:2016hjr}. Left: \emph{transverse} region,
		$\pi/3<|\Delta\phi_Z|<2\pi/3$, right: \emph{towards} region
		$|\Delta\phi_Z|<\pi/3$. }
	\label{fig:ATLAS-Z-jet}
\end{figure}

Before moving on to study rope effects on jets, it is important to
assess whether rope formation drastically changes existing
observables, currently well described by the existing model. In events
with a $\mrm{Z}$-boson present, the most likely place for such a
change to occur, is in the UE. To this end, we use a standard UE
analyses implemented in the Rivet program \cite{Bierlich:2019rhm}.

In \figref{fig:ATLAS-Z-jet}, $N_{\text{ch.}}$ and $\Sigma p_\perp$ for
Z$\rightarrow \ell^- \ell^+ $ events in \pp collisions at 7 TeV are
compared to \atlas data\cite{ATLAS:2014yqy, ATLAS:2016hjr}. The
$\mrm{Z}$-boson is reconstructed from the electron or muon channel
with invariant mass $66 < m_{\ell^- \ell^+} < 166$~GeV in
$|\eta| < 2.5$.

The charged particle multiplicity and summed scalar $p_\perp$
distributions for Z$\rightarrow \mu^- \mu^+$ channel with
$0 < p_\perp^Z < 6$ GeV, are shown in top row of
\figref{fig:ATLAS-Z-jet}. It is seen that adding rope hadronization,
overall preserves the distributions as produced by default \pytppp. We
note that rope hadronization has a slight effect of pushing particles
from lower to higher $\Sigma p_\perp$ regions, which follows from the
$p_\perp$-dependence of the tunnelling probability in
\eqref{eq:tunnel}.

The particle $p_\perp$ in the away region (opposite azimuthal region
to that of the Z boson), balances the $p_\perp^Z$. Hence the towards
and transverse regions with respect to the Z boson are much less
affected by a recoiling jet and therefore have cleaner UE
activity.\footnote{It should here be noted that the charged particle 
  activity in events with a hard interaction such as Z-production is
  generally higher than in minimum bias events.}
These regions are sensitive to the hadronization mechanism, rope
hadronization effects will be apparent here. So we look at the
UE-sensitive observables such as scalar summed
$p_\perp/\delta\eta\delta\phi$ distributions for charged particles in
events with $p_\perp^Z$ in the range 10-20 GeV in the bottom row of
\figref{fig:ATLAS-Z-jet}.
These plots show the $\Sigma p_\perp$
distributions in the transverse $(\pi/3<|\Delta\phi_Z|<2\pi/3)$ and towards 
$(|\Delta\phi_Z|<\pi/3)$ regions\cite{ATLAS:2016hjr}.  We see that the
rope hadronization curve follows the default \pytppp curve, again
preserving the overall physics behaviour of \pytppp, except for a
slight shift in $\Sigma p_\perp$, as in the top right plot.

We conclude that UE measurements are equally well described with
rope hadronization as without, and it is therefore not necessary to re-tune
fragmentation parameters before proceeding to give predictions for jet
observables.

\begin{figure}[t]
	\centering
	\begin{minipage}{1.0\textwidth}
		\includegraphics[width=0.49\textwidth]{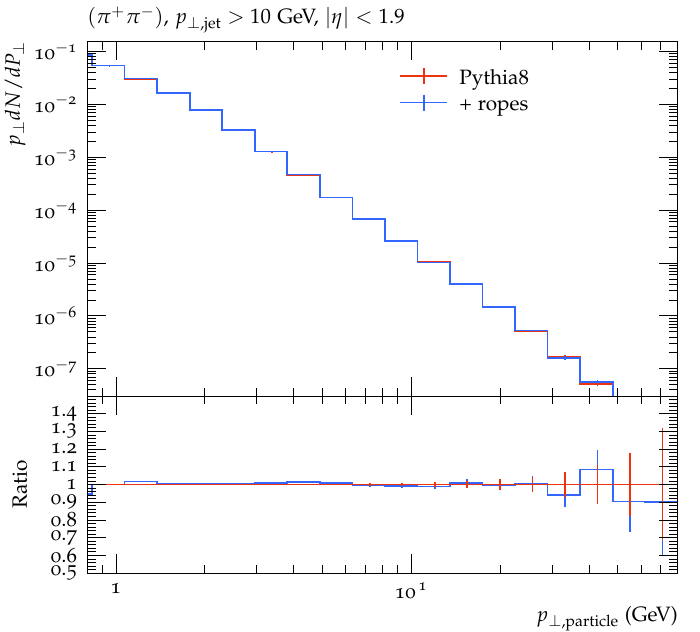}
		\includegraphics[width=0.49\textwidth]{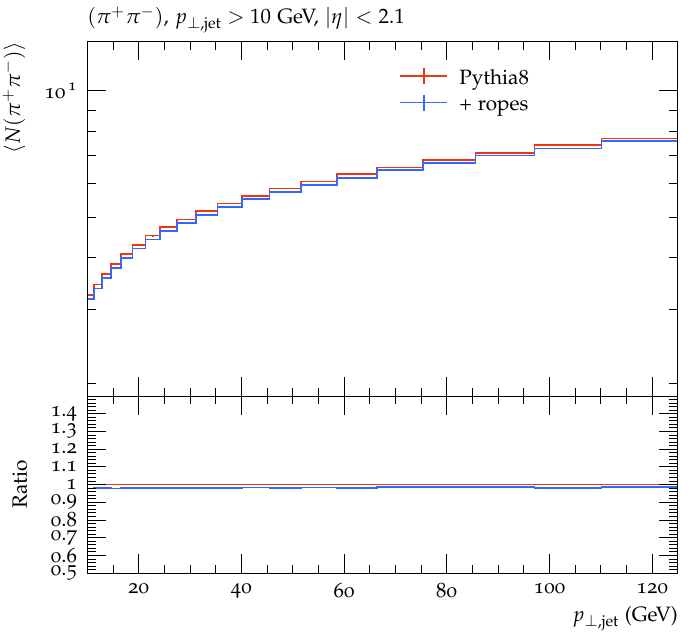}
		\includegraphics[width=0.49\textwidth]{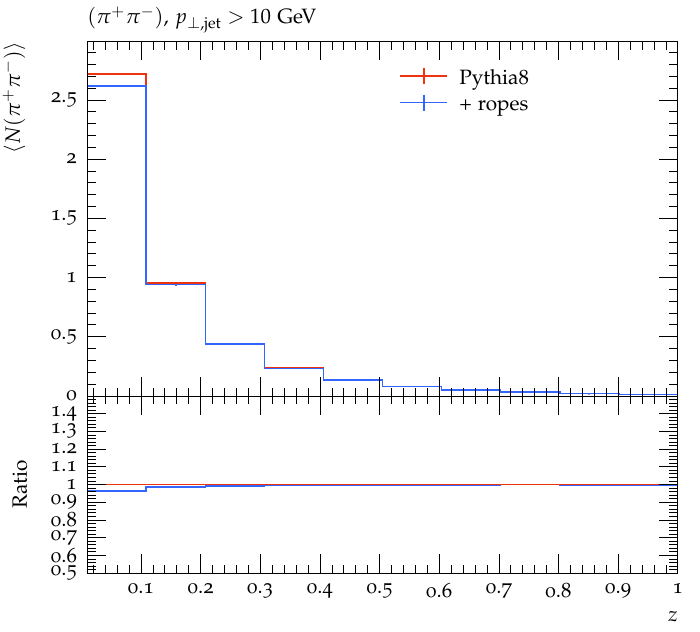}
		\includegraphics[width=0.49\textwidth]{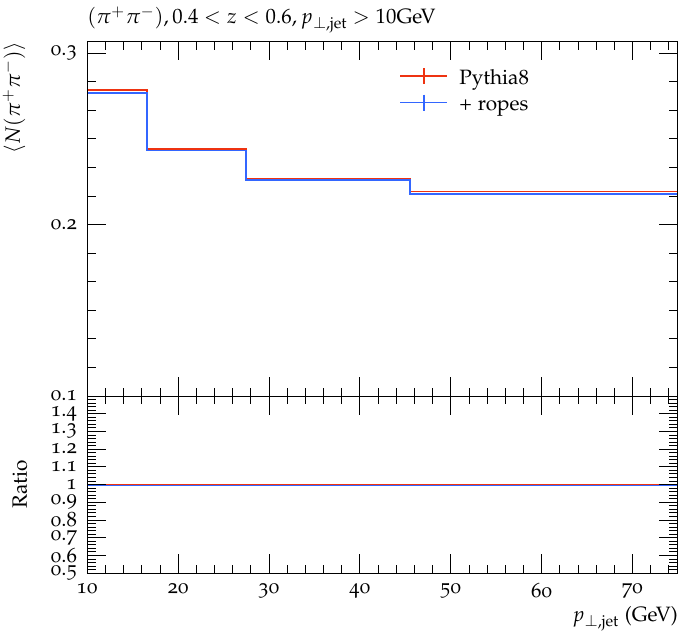}
	\end{minipage}
	\caption{Pion yields in Z$+$jet events in 13 TeV \pp
          collisions vs.\ $p_{\perp, \text{particle}}$ in the UE (top
          left), vs.\ $p_{\perp, \text{jet}}$ in the jet cone (top
          right), as a function of
          $z=p_{\perp, \text{particle}}/p_{\perp, \text{jet}}$ (bottom left), and
          vs.\ $p_{\perp, \text{jet}}$ for $0.4 < z < 0.6$ (bottom right).}
	\label{fig:pion-jet-UE}
\end{figure}

\subsection{Strangeness yields in Z$+$jet events}
\label{sec:strangejet-results-pp}

To investigate experimentally observable consequences of our rope
model in terms of the yield of different hadron species inside jets,
we have chosen to study its effects in Z$+$jets events at LHC
energies. It has been shown in, \eg, \citeref{CMS:2011qzf}, that such
events are very useful for separating regions of phase space dominated
by the UE from the regions dominated by jets. By selecting events
where the Z boson is well balanced by a hard jet in the opposite
azimuthal region, we can study the UE in a cone around the Z, where
there should be very little activity related to the jet, and thus we
can get a good estimate of the UE activity on an event-by-event
basis. In this way we can get a reliable way of correcting jet
observables for UE effects, not only for the transverse momentum of
the jet but also for the flavour content.

\subsubsection{Overall jet features}
\label{sec:overall-jet}

To observe the modification in the flavour production in the jet, 
we want to look at the yield ratios of different hadron
species. Hence we have written a Rivet analysis where we first locate a reconstructed Z boson for
$m_{\mu^- \mu^+}$ in the range 80-100 GeV and $|\eta| < $ 2.5 and
search for the hardest associated jet in the opposite azimuthal
hemisphere. We further restrict the Z boson by requiring it to be
within $|\eta| <$ 1.9 and $p_\perp^Z> 8$~GeV using the standard
Z-finding projection in \rivet. Once we find such a Z boson in the
event, we search for the associated hardest (charged particle) jet
using the anti-k$_T$ \cite{Cacciari:2008gp} algorithm with a radius
$R_j=0.4$ in $|\eta| < 2.1$ with the azimuthal separation
$\Delta \phi_{\text{jet},Z} \geq 2\pi/3$.

\begin{figure}[t]
  \centering
  \includegraphics[width=0.75\textwidth,height=0.45\textheight]{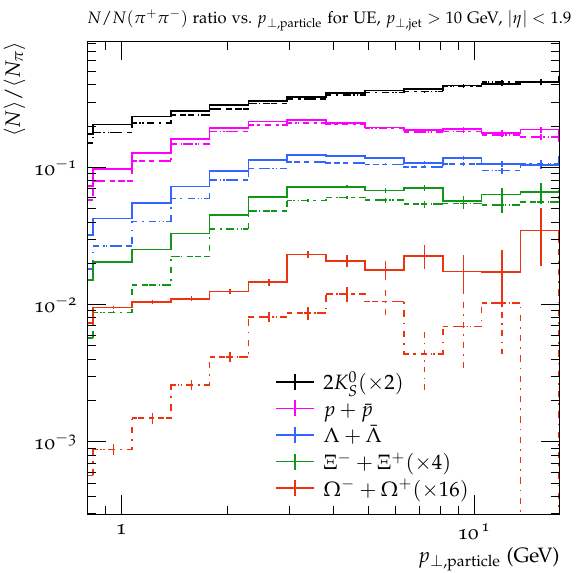}
  \caption{Yield ratio of different strange hadron species and protons
    to pions in the UE cone vs.\ $p_{\perp, \text{particle}}$, scaled
    by factors to show them clearly. Solid lines are with rope
    hadronization and dot-dashed lines are for default \pytppp.}
  \label{fig:hadronyield-UE}
\end{figure}

To subtract UE contributions from the jet $p_\perp$, we calculate
a characteristic $\Sigma p_{\perp, \text{UE}}$, by summing up the $p_\perp$ of the charged
final state particles (not including muons from the Z decay) that lie
within a cone of radius $\sqrt{2}R_j$ around the Z boson. Therefore, for a given event, the yields of the particles
is calculated twice: once within the jet cone, then within a cone
of radius $\sqrt{2}R_j$ with respect to the Z boson. The latter serves
as our underlying event reference and we subtract half of this yield
from the yield inside the jet cone to get the final yield of the
hadrons in that event associated with the jet. Denoting the initial jet-$p_\perp$ as $p_{\perp,\text{pseudojet}}$, the corrected
$p_{\perp,\text{jet}}$ becomes:

\begin{equation}
\label{eq:jet-pT}
	p_{\perp,\text{jet}} = p_{\perp,\text{pseudojet}} - 0.5 \times \Sigma p_{\perp,\text{UE}}
\end{equation}
and the corresponding yields:
\begin{equation}
\label{eq:yield}
\text{yield}_{\text{jet}} = \text{yield}_{\text{pseudojet}} - 0.5 \times \text{yield}_{\text{UE}}
\end{equation}

This method of UE subtraction can easily be extended to \pA and \AA collisions to
give a comparable result among the three systems. Similar methods have previously
been used in heavy ion collisions \cite{Neufeld:2012df}. We do this analysis for \pp collisions at $\sqrt{s}=13$ TeV with $p_{\perp, \text{jet}} \geq$ 10 GeV for string radius $R = 1$~fm. 

\begin{figure}[t]
  \centering
  \includegraphics[width=0.75\textwidth,height=0.45\textheight]{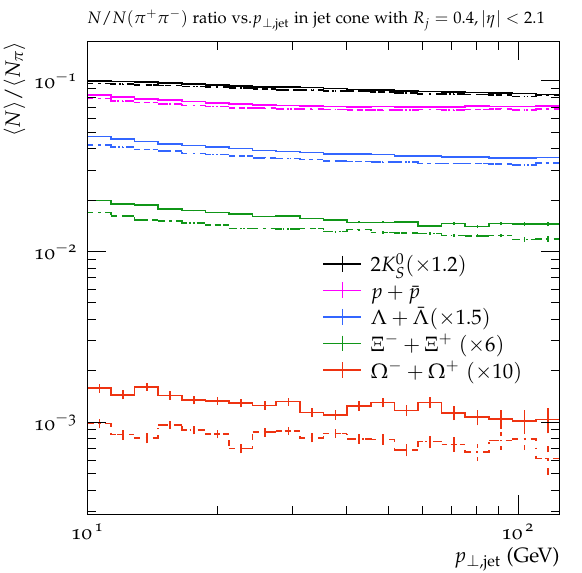}
  \caption{Yield ratio of different strange hadrons and protons to
    pions in the jet cone, for $R_j=0.4$ vs.\ $p_{\perp, \text{jet}}$,
    scaled by factors to show them clearly. Solid lines are with rope
    hadronization and dot-dashed lines are for default \pytppp.}
  \label{fig:hadronyield-jet}
\end{figure}

To examine the model performance in reproducing general features of
the jets, such as particle multiplicity as a function of their
transverse momentum and of the transverse momentum of the jets, we
look at the pions. Our rope model is known to have very small effects
on the overall multiplicity \cite{Bierlich:2014xba}, and we know that
pions in general are dominating the particle production, even though
we expect a slight drop in pions, since high $\keff$ will favour
strange hadrons and baryons over pions. In \figref{fig:pion-jet-UE} we
show the pion yield as a function of particle $p_\perp$ in the UE cone,
and the UE-subtracted yield as a function of $p_{\perp, \text{jet}}$
in the jet cone in \figref{fig:pion-jet-UE}. We also show the pion
yield with respect to
$z=p_{\perp,\text{particle}}/p_{\perp,\text{jet}}$ and in the mid-$z$
region as a function of $p_{\perp,\text{jet}}$. Indeed we find that
the rope effects are very small for pion production, both in the UE
and in the jet, with the possible exception of the lowest bin in the
$z$ distribution.
We will revisit the bottom row plots in connection with strangeness
yields in the jet cone in section \ref{sec:subjet-results}.

In the UE region, the density of strings is high resulting in a higher number of overlaps among them. As a result, we would expect large effects due to rope hadronization in the UE. In order to observe this effect, we look at the yield ratio of the strange hadrons to pions
in the UE cone. In \figref{fig:hadronyield-UE}, we show the yield
ratio to pions for strange mesons ($K^0_S$) and baryons ($\Lambda$,
$\Xi$ and $\Omega$) and protons with respect to
$p_{\perp,\text{particle}}$. Yields of $\Xi$ and $\Omega$ baryons have
been scaled by a multiplicative factor to show them in comparison to
the other species. As expected, the different yields are higher with
rope hadronization turned on as compared to default \pytppp. The
highest enhancement for each species is observed for the lowest
$ p_{\perp, \text{particle}}$ ranges which subsequently decreases for
higher particle $p_\perp$ (which follows \figref{fig:kappa-mult-pT} in
section \ref{sec:model-results}). Therefore, this plot show us that
with rope hadronization, we get increased yields of baryons and
strangeness.  This plot also shows us the UE contribution to
strangeness yields to that of within the jet.

Turning to flavour production \textit{inside} the jet cone in \figref{fig:hadronyield-jet}, we show the UE-subtracted yield ratio to pions for the same set of hadron species as before, now with respect to $p_{\perp,\text{jet}}$. As rope hadronization will increase both strangeness and baryon production, the largest enhancement is expected for multistrange baryons. For $K^0_S$, only a slight increase is observed, while the increase for protons is higher. The $\Lambda$ yield due to rope hadronization is even higher due to combined baryon and strangeness enhancement. The yield of $\Xi$ is $\sim$ 20\% higher due to rope hadronization, than default \pytppp and the $\Omega$ yield with rope hadronization is more than 50\% higher. This shows that both baryon and strangeness yields are enhanced by rope hadronization.
We note that the increase in the yield ratio due to rope hadronization is rather constant over all $p_{\perp, \text{jet}}$. Hence if we look at the enhancement as a function of the transverse momentum ratio of the particle species to that of the jet, that would help us identify the $p_\perp$ ranges where rope effects are higher. 

\begin{figure}[t]
  \centering
  \begin{minipage}{1.0\textwidth}
    \begin{center}
      \includegraphics[width=0.49\textwidth]{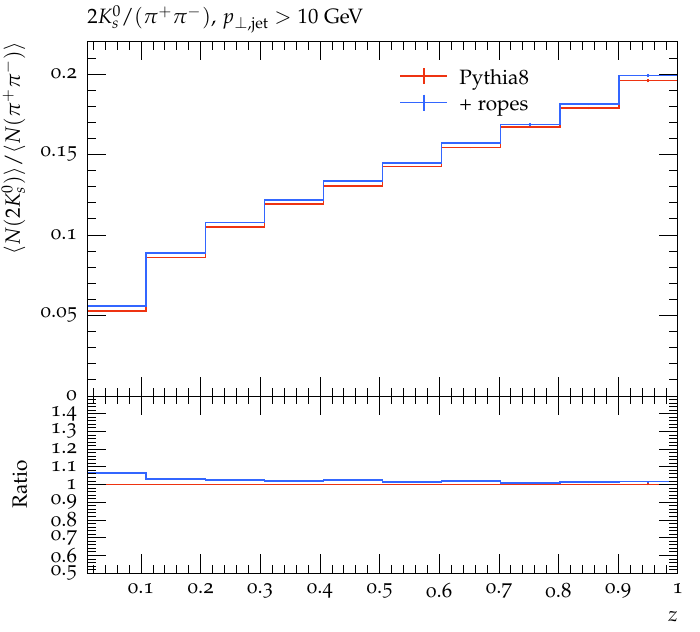}
      \includegraphics[width=0.49\textwidth]{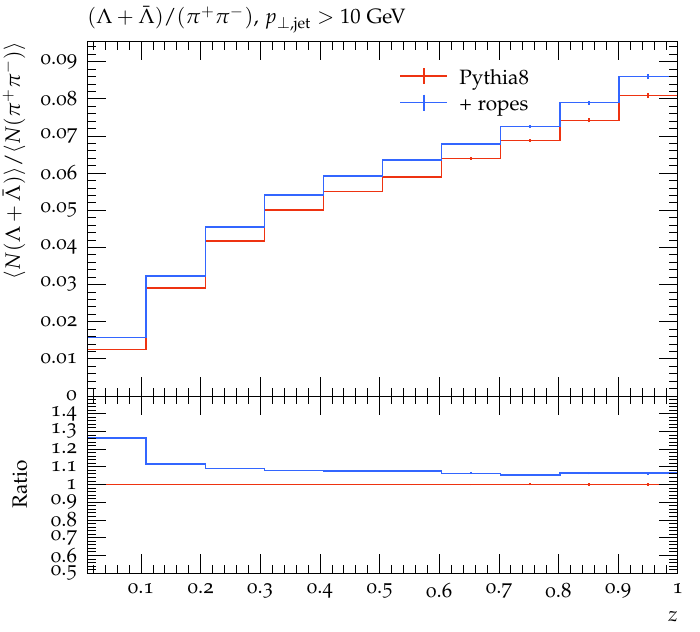}
      \includegraphics[width=0.49\textwidth]{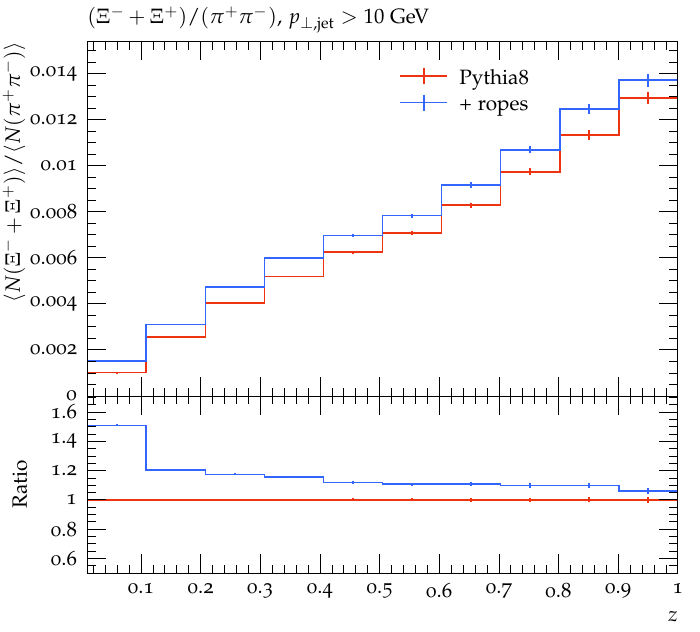}
      \includegraphics[width=0.49\textwidth]{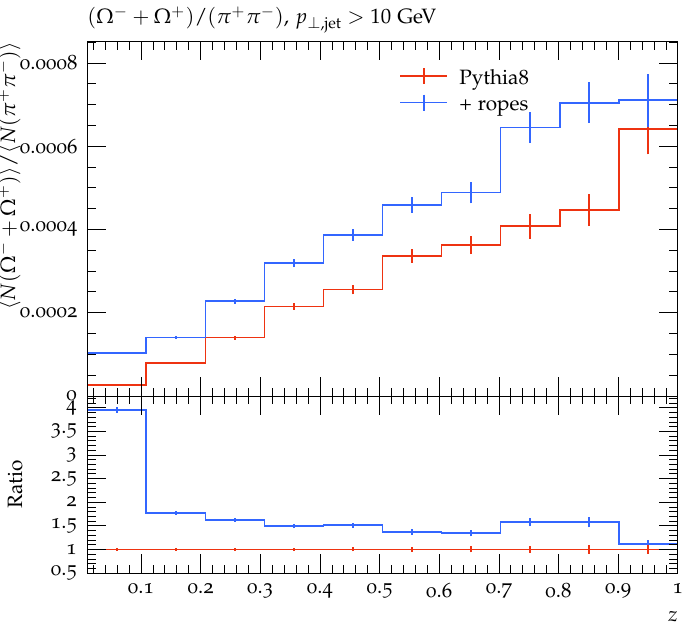}
    \end{center}
  \end{minipage}
  \caption{Yield ratios as a function of
    $z=p_{\perp, \text{particle}}/p_{\perp, \text{jet}}$ for \pp
    collisions at $\sqrt{s}$=13 TeV: $2K^0_S/(\pi^+ \pi^-)$ (top left)
    $(\Lambda +\bar{\Lambda})/(\pi^+ \pi^-)$ (top right),
    $(\Omega^-+\Omega^+)/(\pi^+ \pi^-)$ and
    $(\Xi^-+\Xi^+)/(\pi^+ \pi^-)$ (bottom left).}
  \label{fig:particle-jet-pT}
\end{figure}

\subsubsection{Jet substructure observables}
\label{sec:subjet-results}

Now we take a closer look at the particle to pion yield ratios as a function of $z$ and $p_{\perp, \text{jet}}$. Studies have been performed where the ratio of $p_\perp$ of the individual sub-jets to that of the leading jet serves as a distinguishing observable for jet modification\cite{Apolinario:2017qay}. Since we want to look at the strange flavour yields in the jet cone, we take a simpler approach. We only plot the yield ratios in bins of $z$, which is the ratio of the particle $p_\perp$ to the jet $p_\perp$. 

In \figref{fig:particle-jet-pT}, we show the yield ratio of strange hadrons to pions \textit{vs.}\ $z$.
We observe that the particle yields are increased at low (close to the UE) to intermediate $z$ values. Furthermore, this enhancement is smaller for $K^0_S$ and larger for the strange baryon $\Lambda$, and for multistrange baryons $\Xi$  and $\Omega$ as expected. However, strangeness and baryon enhancement drops at higher $z$. This highlights the
behaviour that rope hadronization effects decrease with higher $p_\perp$, as we noted
in \figref{fig:kappa-mult-pT} in section \ref{sec:model-results}.
\begin{figure}[t]
  \centering
  \begin{minipage}{1.0\textwidth}
    \begin{center}
      \includegraphics[width=0.49\textwidth]{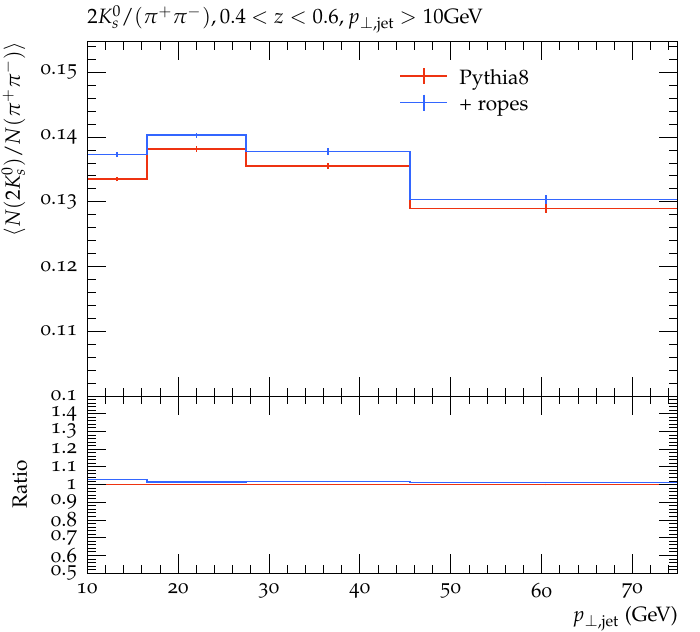}
      \includegraphics[width=0.49\textwidth]{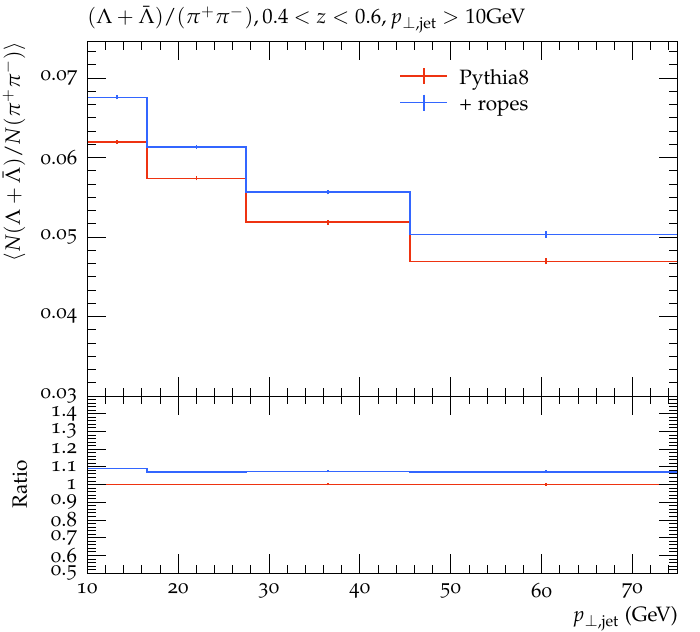}
      \includegraphics[width=0.49\textwidth]{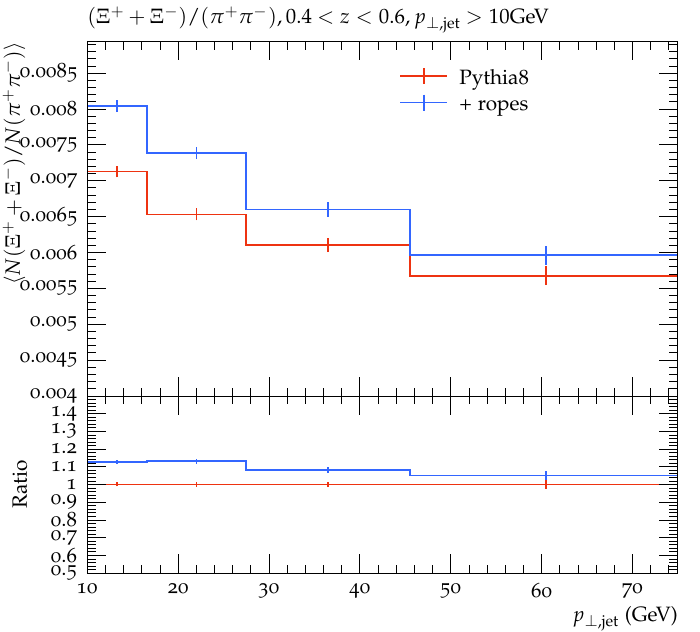}
      \includegraphics[width=0.49\textwidth]{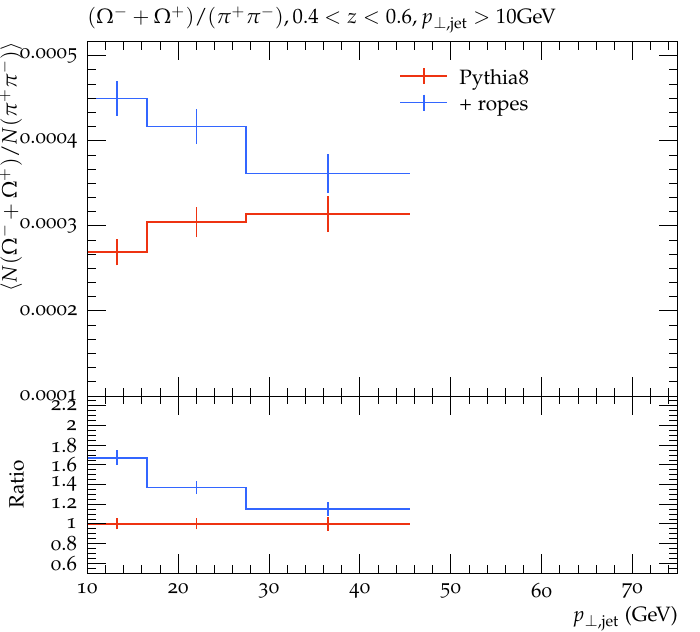}
    \end{center}
  \end{minipage}
  \caption{Yield ratios of particles with $0.4<z<0.6$, as a function of \ $p_{\perp,
      \text{jet}}$ for \pp collisions at $\sqrt{s}$=13 TeV:
    $2 K^0_s/(\pi^+ \pi^-)$ (top left),
    $(\Lambda +\bar{\Lambda})/(\pi^+ \pi^-)$ (top right),
    $(\Xi^-+\Xi^+)/(\pi^+ \pi^-)$ (bottom left) and
    $(\Omega^-+\Omega^+)/(\pi^+ \pi^-)$ (bottom right).}
  \label{fig:zcut-jet-pT}
\end{figure}

We note that, even though the parallel frame formalism allows the calculation of \keff~in events with jets, the current implementation is lacking in the region $z \approx 1$, as already mentioned in \sectref{sec:monte-carlo-impl}. The previously mentioned \textit{Catch-22} situation, is purely related to the implementation, and can be further understood by considering the shape of the Lund symmetric fragmentation function in \eqref{eq:lu-frag}, which is vanishing near $z = 1$. For a particle with $z$ close to one, the pre-sampled overlap is therefore likely to have been calculated with a too-small $z$, which in turn means that it is calculated for the wrong part of the string. In \pp collisions this effect is small but non-negligible, which we have confirmed by an \textit{a posteriori} check (as the correct overlaps can be calculated after the fact, but too late to be used in event generation). Another issue, which would be present even in a perfect implementation, and therefore potentially more severe, is the absence of interactions between hadrons formed early in time, and their surrounding environment. For most of the produced particles, and in particular in \pp, this effect should also be small. But in the case of high $z$, the particle is always produced early, and the effect could be larger. We plan to develop the model further in this direction, but in the meantime we will in the following show results for particles at intermediate $z$ values ($0.4 < z < 0.6$) where the effects arising from both these issues, should be negligible.

To test the modification in flavour yields at mid-$z$ values, we look at particle yields as a function of $p_{\perp, \text{jet}}$. Since these particles are neither close to the tip of the jet, nor to the UE, it is more reasonable to the trial-hadron sampling of $\keff$ in these regions. Moreover, as the jet $p_\perp$ increases, the particles get further and further away from the UE.
In \figref{fig:zcut-jet-pT}, we show the yield ratio of strange
hadrons to pions in the 0.4 < $z$ < 0.6 region vs.\ $p_{\perp, \text{jet}}$. We observe that the yields from the rope hadronization case are distinct compared to default \pytppp. The individual strange hadron yield to pion yield ratio increases as we go from the $K^0_s$ meson to the $\Lambda$ baryon (top row plots). For multistrange baryons, $\Xi$ and $\Omega^-$ (bottom row plots), rope effects are amplified due to higher number of strange quarks, resulting in a 20\% - 50\% increase in their yields in low $p_{\perp, \text{jet}}$ ranges.  However, as mentioned before, we would expect the enhancement in the yields to drop at higher $p_{\perp, \text{jet}}$ bins. This effect is rather small for $\Lambda$
but prominent for $\Xi$ and $\Omega$. $\Omega$ (bottom right plot) is only shown up to 45 GeV due to statistics.

\section{Conclusion}
\label{sec:conclusion}

We have here presented a study on how an effect from a dense system
of colour fluxtubes might be observed as strangeness enhancement in
jets in high multiplicity pp events. In such events it is essential to
properly estimate the interaction between non-parallel strings,
including strings connected to a hard scattered parton and strings in
the underlying event. This problem was solved in
\citeref{Bierlich:2020naj}, where the interaction of all string pairs
can be calculated in a Lorentz frame, where two string pieces lie
symmetrically in two parallel planes. We here show results for
jet-triggered high-multiplicity \pp collisions. The generalization to
\pA and \AA collisions (using the \angantyr model \cite{Bierlich:2018xfw}) 
will be presented in a future publication.

The interacting strings can form ``colour ropes'', which hadronize in a
stepwise manner by $q\bar{q}$ pair creation. The increased energy in
the rope gives a higher "effective string tension", $\keff$, which
increases the number of strange quarks and diquarks in the
breakups. In \sectref{sec:model-results} we found that this results in
an increase of $\keff$ with multiplicity in \pp events at LHC
energies. It is interesting to note that the increase for a given
multiplicity is almost independent of the collision energy.

As expected we also found that the increase is quite dependent on the
transverse momentum, since high-$p_\perp$ particles are typically
produced in jets where the strings are not parallel with the bulk of
the strings in the underlying event, thus reducing the effective
overlap with these. The important question is then if the rope model,
despite being reduced in jets, anyway will result in a modification of
the hadron composition of jets.

To study the effects on jets we focused our investigation on Z+jet
events, with the Z decaying to lepton pairs. As pointed out in \eg\
\citeref{CMS:2011qzf}, it is possible, in such events, to get a
relatively clean separation between the jets and the particle
production in the underlying event. In particular the hadrons produced
in a cone around the direction of the Z particle should have very
little to do with the recoiling jet, and can therefore be used to
correct any observable in the jet cone for underlying-event
contributions on an event-by-event basis.

The modified $\keff$ also affects the fragmentation parameters. In 
\sectref{sec:jet-prop-results} results
for multiplicity and the transverse momentum distribution in the
underlying event in pp Z+jet events, were compared with results from default \pytppp
and with data from ATLAS. After confirming that the rope hadronization
gives negligible effects on these general features of the underlying event, we
feel comfortable that we can study strangeness and baryon enhancement
in the jets in a way, which is not biased by the underlying-event
corrections.

In \sectref{sec:strangejet-results-pp} our main
results for strangeness and baryon number enhancement in jets were 
presented, with the underlying event subtracted. We note that the effect is most
important for strange baryons, and growing with the number of strange
quarks.  Thus it is largest for $\Omega$ baryons, and from the plots
showing the $\Omega/\pi$ ratio as a function of the jet transverse
momentum, we note that rope effects are very small for large jet
$p_\perp$ as expected, but quite noticeable for low jet $p_\perp$.

From this we conclude that it may indeed be possible to find jet
modifications due to collective effects, in our rope model, in small
collision systems. The size of the effect is, however, a bit
uncertain. In part this is due the uncertainty in the transverse size
of the string, and our canonical choice of $R=1$~fm may be a bit
large. Although it should be possible to tune this parameter to fit
the overall strangeness and baryon enhancement, it is then also
important to also take into account the effects of repulsion between
the strings. Both of these effects will be addressed in future
publications.

Looking ahead, it is also interesting to investigate the effects of
colour reconnection, in particular models that include junction
formations, which will also influence the baryon production. In the
end we hope to develop a picture where most collective effects can be
interpreted as interactions among strings, not only in \pp\ collisions
but also in \pA\ and \AA.

\section*{Acknowledgements}
This work was funded in part by the Knut and Alice Wallenberg
foundation, contract number 2017.0036, Swedish Research Council, contracts
number 2016-03291, 2016-05996 and 2017-0034, in part by the European
Research Council (ERC) under the European Union’s Horizon 2020
research and innovation programme, grant agreement No 668679, and in
part by the MCnetITN3 H2020 Marie Curie Initial Training Network,
contract 722104.

\appendix

\section{Dependence of fragmentation parameters on \boldmath\keff}
\label{sec:kappaeffects}

There are several hadronization parameters in \pytppp, and even if
they are in principle independent, several of them has an implicit
dependence on the string tension. In our implementation of the rope
hadronization, we take the parameters as tuned to \ee\ data, where we
expect no rope effects, and for each breakup in the string
fragmentation we rescale the parameters according to the estimated
change in string tension at that point, due to the presence of
overlapping string fields. The parameters under consideration is the
same as in our previous implementation \cite{Bierlich:2014xba}, and
the dependence of the string tension is also the same. For
completeness we list them here, but for further details we refer to
\cite{Bierlich:2014xba}.

In the following we will denote the change in string tension by $h$,
according to $\kappa\mapsto\keff=h\kappa$. The following parameters is
affected:
\begin{itemize}
\item $\rho$ (\texttt{StringFlav:probStoUD}\footnote{This is the
    parameter name in \pytppp.}): the suppression of $s$ quark
  production relative to $u$ or $d$ type production. This parameter
  has a simple scaling
  \begin{equation}
    \label{eq:scale-rho}
    \rho\mapsto\tilde{\rho} = \rho^{1/h}.
  \end{equation}
\item $x$ (\texttt{StringFlav:probSQtoQQ}): the suppression of
  diquarks with strange quark content relative to diquarks without
  strange quarks (in addition to the factor $\rho$ for each extra
  $s$-quark) also scales like
  \begin{equation}
    \label{eq:scale-x}
    x \mapsto \tilde{x} = x^{1/h}.
  \end{equation}
\item $y$ (\texttt{StringFlav:probQQ1toQQ0}): the suppression of spin
  1 diquarks relative to spin 0 diquarks (not counting a factor three
  due to the number of spin states of spin 1 diquarks) again scales like
  \begin{equation}
    \label{eq:scale-y}
    y \mapsto \tilde{y} = y^{1/h}.
  \end{equation}
\item $\sigma$ (\texttt{StringPT:sigma}): the width of the transverse
  momentum distribution in string break-ups. This is directly
  proportional to $\sqrt{\kappa}$, giving
  \begin{equation}
    \label{eq:2}
    \sigma \mapsto \tilde{\sigma} = \sigma\sqrt{h}.
  \end{equation}
\item $\xi$ (\texttt{StringFlav:probQQtoQ}): the global probability of
  having a diquark break-up relative to a simple quark break-up. This
  has a somewhat more complicated $\kappa$ dependence and also has
  uncertainties related to the so-called popcorn model as described in
  \cite{Bierlich:2014xba}. We decompose it as three different
  parameters, $\xi=\alpha\beta\gamma$ with different
  $\kappa$-dependence, where $\beta$ is related to the probability to
  have a $q\bar{q}$ fluctuation in general in the popcorn model which
  is independent of $\kappa$ and is treated as an independent
  parameter, while $\gamma$ is related to the masses and scales as
  \begin{equation}
    \label{eq:scale-gamma}
    \gamma \mapsto \tilde{\gamma} = \gamma^{1/h},
  \end{equation}
 and $\alpha$ is related to the different di-quark states with an
  indirect dependence on $\rho$, $x$, and $y$
  \begin{equation}
    \label{eq:scale-alpha}
    \alpha \mapsto \tilde{\alpha}= 
    \frac{1 + 2\tilde{x}\tilde{\rho} + 9\tilde{y} + 6\tilde{x}\tilde{\rho} y
      + 3\tilde{y}\tilde{x}^2\tilde{\rho}^2}
    {2+\tilde{\rho}}.
  \end{equation}
  Taken together we get the following dependence:
  \begin{equation}
    \label{eq:scale-xi}
    \xi=\alpha\beta\gamma \mapsto \tilde{\xi} =
    \tilde{\alpha}\beta\left(\frac{\xi}{\alpha\beta}\right)^{1/h}.    
  \end{equation}
\item $b$ (\texttt{StringZ:bLund}): the parameter in the symmetric
  fragmentation function \eqref{eq:lu-frag} scales with the
  $\rho$-parameter as follows
  \begin{equation}
    \label{eq:scale-b}
    b \mapsto \tilde{b} = \frac{2 + \tilde{\rho}}{2 + \rho}\,b.
  \end{equation}
\item $a$ (\texttt{StringZ:aLund}): the other parameter in
  \eqref{eq:lu-frag} has an indirect dependence on $b$ through the
  normalisation of the splitting function, $f(z)$. Keeping the
  normalisation unchanged does not give a simple analytic form for the
  scaling of $a\mapsto\tilde{a}$, and instead we use a numeric
  integration procedure.\end{itemize}

\bibliography{bibliography}

\nolinenumbers

\end{document}